\documentclass[fleqn,12pt,twoside]{article}
\usepackage{espcrc1}
\usepackage{times}

\usepackage{graphicx}
\usepackage[figuresright]{rotating}

\newcommand{\AmS}{{\protect\the\textfont2
  A\kern-.1667em\lower.5ex\hbox{M}\kern-.125emS}}
\title{ Why the OZI rule is so strongly violated in $J/\Psi$ decays?}
\author{{ L. Ya. Glozman}\vskip5mm Institute for Theoretical
Physics, University of Graz, Universit\"atsplatz 5, A-8010
Graz, Austria\footnote{e-mail: leonid.glozman@uni-graz.at}}

\begin{document}
\maketitle


\begin{abstract} 
The new $f_0(1790)$ meson recently observed by BES collaboration in
$J/\Psi$-decay, is seen only in the OZI-forbidden channel. It is
shown that  chiral symmetry restoration in excited hadrons
implies a new selection rule of dynamical origin
that forbids some of the
OZI-favoured mechanisms of decays. Hence decays  into channels
that are suppressed by OZI can become dominant.

\end{abstract}
\bigskip

\bigskip

A recent remarkable result by BES collaboration on $J/\Psi$
decay into $\phi \pi^+ \pi^-$ and  $\phi K^+ K^-$ has established
a new scalar meson, $f_0(1790)$, with $M=1790^{+40}_{-30}$ MeV and
$\Gamma = 270^{+60}_{-30}$ MeV with the dominant 
$\bar n n = \frac{\bar u u + \bar d d}{\sqrt 2}$ valence quark
content \cite{BES}. This state is observed as a $15\sigma$ peak
in $\pi^+ \pi^-$ while its decay into $K K$ is strongly suppressed.
Actually this state was seen earlier in the London-S.Petersburg
partial wave analysis of $\bar p p$ at LEAR at $1770 \pm 12$ MeV 
\cite{BUGG1,BUGG2} and some other reactions \cite{BUGG2}.
\footnote{The spin 0 assignement of this state is preferred
over spin 2 in the BES publication, while such an assignement is
conclusive in $\bar p p$ \cite{BUGG1,BUGG2}.}
It was also claimed in those publications that this state must be
a $\bar n n$ state and should not be confused with the nearby
$f_0(1710)$, which is pedominantly a $\bar s s$ state. This claim
of the London-S.Petersburg group has been indeed confirmed by BES
since $f_0(1710)$ is clearly seen in $J/\Psi \rightarrow \omega (K^+K^-)$
and $J/\Psi \rightarrow \phi(K^+K^-)$, while $f_0(1790)$ is not visible
there. There is definite peak in $J/\Psi \rightarrow \phi(\pi^+\pi^-)$
for   $f_0(1790) \rightarrow \pi^+\pi^-$ while the signal for
$f_0(1710)$ is very strongly suppressed.

This $f_0(1790)$ state is very interesting as it rules out
many existing scenarios for the light mesons. Indeed, there is a well
established state with the predominantly $\bar n n$ content $f_0(1370)$
and the new discovered  $f_0(1790)$ must be its radial excitation.
In between there are well established $f_0(1500)$ and $f_0(1710)$
which are predominantly glueball and $\bar s s$ states, respectively,
 with some
mixing \cite{CLOSE}. Actually there are further radial excitations
of $\bar n n$ $f_0$ mesons: $f_0(2040)$ and $f_0(2337)$ which are seen
in a few different channels in $\bar p p$ \cite{BUGG3,BUGG2}.

A very intriguing observation of BES is that the $f_0(1790)$ state
is seen only in the OZI-forbidden channel
$J/\Psi \rightarrow \phi f_0(1790) \rightarrow \phi \pi^+\pi^-$.
Indeed, the OZI rule would require that the $\bar n n$ state should be
accompanied by $\omega$, but not by $\phi$, see Fig. 1, because
$\phi$ has a well established $\bar s s$ content. However, $f_0(1790)$
 is not seen in $J/\Psi  \rightarrow \omega (\pi^+\pi^-)$ and
is seen only in $J/\Psi  \rightarrow \phi(\pi^+\pi^-)$, see Fig. 2.
Then there arise two independent questions. (i) Why the OZI
mechanism does not work? (ii) Which specific mechanism
is behind the Fig. 2 and why then the OZI forbidden  doubly 
disconnected  mechanism of decay
$J/\Psi \rightarrow \omega f_0(\bar n n)$ (with the $\phi$-meson
on Fig. 2 to be substituted by the $\omega$-meson) is 
relatively suppressed?

A purpose of this note is to demonstrate that  restoration of
chiral symmetry in excited hadrons
 \cite{G1,CG,G2,G3,G4,G5,Beane,Swanson,Shifman,R,ECT} 
would imply  a new dynamical selection rule that forbids
$f_0(1790)$ to be accompanied by $\omega$ meson in $J/\Psi$ decay
via the OZI-allowed mechanism.
Then  this decay  via the OZI-forbidden
diagram(s) like of Fig. 2 can be indeed dominant.
We do not attempt, however, to answer the interesting question
(ii). Clearly this answer will necessarily invoke 
microscopical models for decay, which is beyond the scope of
this paper.
The suppression of the OZI-favoured mechanism by chiral symmetry
  is a general claim and equally applies to other possible decays of the heavy
quarkonium into two mesons with one of the mesons being in the
chirally restored regime.

\begin{figure}
\begin{center}
\includegraphics*[width=10cm]{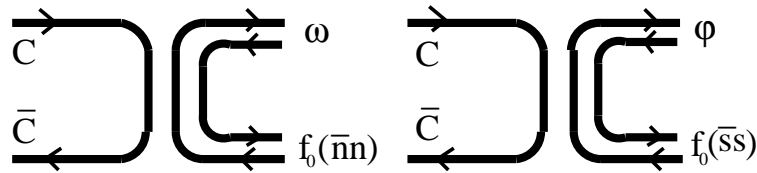}
\end{center}
\caption{The OZI-favoured decays into two mesons.}
 
\end{figure}

\begin{figure}
\begin{center}
\includegraphics*[width=10cm]{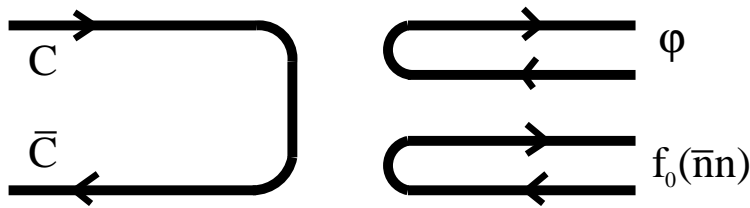}
\end{center}
\caption{The OZI-forbidden decay.}
 
\end{figure}

Restoration of the chiral symmetry in excited hadrons requires that
the highly excited hadrons in the $u,d$ sector fall into representations
of $SU(2)_L \times SU(2)_R$ or $U(1)_A$ group. There are 
empirical evidences for this phenomenon both in excited baryons and
mesons \cite{G1,CG,G2,G3,G4,G5}, for overview see Ref.
\cite{R}. In particular, a spectrum of 
excited $\pi$ and 
$\bar n n$ $f_0$ mesons represents one of the clearest manifestations
from this phenomenon \cite{G2}, see Fig. 3.
 Indeed, if chiral symmetry was not broken
in the QCD vacuum, then all $\pi$ and $\bar n n$ $f_0$ mesons would
be systematically degenerate and fall into (1/2,1/2) representations
of $SU(2)_L \times SU(2)_R$. Spontaneous breaking of chiral symmetry
in the vacuum removes this degeneracy and the ground state pion
turnes into Goldstone mode with zero mass. However, the
quark condensates of the
vacuum become progressively irrelevant in the radially excited hadrons
where the valence quarks decouple from the quark condensates and
the chiral symmetry gets approximately restored in the given hadronic state
and in the spectrum  \cite{CG,Shifman}. Fundamentally this happens because quantum
fluctuations of the quark fields, which are the origin of spontaneous
breaking of chiral symmetry, get suppressed in excited hadrons and
a semiclassical description becomes adequate \cite{G5}.
In particular, an approximate degeneracy within the
$\pi(1800) - f_0(1790)$ chiral pair demonstrates how insignificant are
the remaining effects of chiral symmetry breaking in the vacuum
for the $n=3$ state where $n$ is the radial quantum number.

While for the large $N_c$ limit the effective restoration
of chiral symmetry in excited mesons has been proven \cite{CG,Shifman},
this phenomenon should be still considered as a conjecture
in the $N_c=3$ world until
the missing states in the chiral multiplets will have been found.
One can try to argue that the approximate parity doublets seen
in the spectrum are accidental and the states of opposite
parity actually belong to close shells with alternating parity in
the spirit of the naive constituent quark model. However, the assumptions
of the constituient quark model, if valid for the $N_c=3$ world,
must  be equally valid for the large $N_c$ mesons. In the latter case 
the chiral restoration does occur and hence the naive constituent quark 
picture of excited mesons is ruled out, at least in the large $N_c$ limit.

Now comes the key point. Once the chiral symmetry is restored,
then the valence quark wave function is fixed in terms of the
right- and left-handed quark components:

$$ f_0:   \frac{\bar R L + \bar L R}{\sqrt 2},$$

$$ \pi:   \frac{\bar R \vec \tau L - \bar L \vec \tau R}{\sqrt 2},$$

\noindent
where $\tau$ are isospin Pauli matrices and
$L$ is a column consisting of the left-handed $u$ and $d$
quarks, while $R$  is a column consisting of the right-handed $u$ and $d$
quarks. Given almost perfect degeneracy of $f_0(1790)$ and
$\pi(1800)$ states one naturally assumes that an admixture
of the components violating chiral symmetry  in $f_0(1790)$
is very small. Then
the OZI-favoured decay  mechanism via two or three perturbative gluons
(depending on the C-parity of the decaying heavy quarkonium), see Fig. 4,
would require that the same right-left structure should persist
for the other quark-antiquark pair that is a source for the 
accompanied meson. This is because the quark-gluon vertex conserves
 chirality for massless quarks. However, the
$\frac{\bar R L + \bar L R}{\sqrt 2}$ quark pair has positive
parity and cannot be a source (interpolating field)
 for the $\omega$ meson, which is
a vector particle. Then one concludes that the OZI-favoured
decay mechanism  of $J/\Psi  \rightarrow \omega f_0(1790)$ is forbidden
by chiral symmetry. Clearly, there are no chiral symmetry
restrictions for the OZI-forbidden channel of Fig. 2.

\begin{figure}
\begin{center}
\includegraphics*[width=7cm,angle=-90]{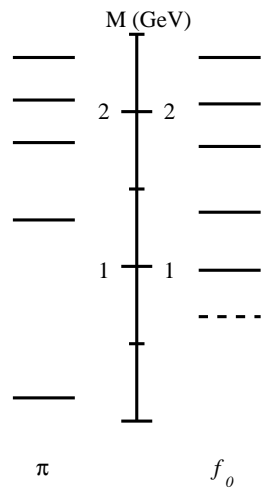}
\end{center}
\caption{Pion and $n \bar n$ $f_0$ spectra.}
\end{figure}

It is rather obvious that this and similar chiral symmetry
selection rules of dynamical origin 
are rather general and can (and should) be
observed in many other cases. For example, the same rule
is directly relevant to $f_0(1370)$, that is also observed in the OZI-forbidden
decay, $J/\Psi  \rightarrow \phi f_0(1370)$, and is not seen in the
 $J/\Psi  \rightarrow \omega f_0(1370)$,
which is another claim of BES. Indeed, as is seen from the Fig. 3,
 $\pi(1300 \pm 100) - f_0(1370^{+130}_{-170})$ is the lowest-lying
approximate chiral pair seen in the $\pi$-$f_0$ spectra.

In general, in any two-meson decay of the heavy quarkonium and
once one of the mesons is in the chirally restored regime the chiral
symmetry does not support generically the OZI decay mechanism\footnote{
Except for the cases when both mesons have equal quantum numbers.}
 and
hence the strong violations of the OZI rule should be expected. This
can be studied at BES, CLEO  and other similar facilities.

\begin{figure}
\begin{center}
\includegraphics*[width=10cm]{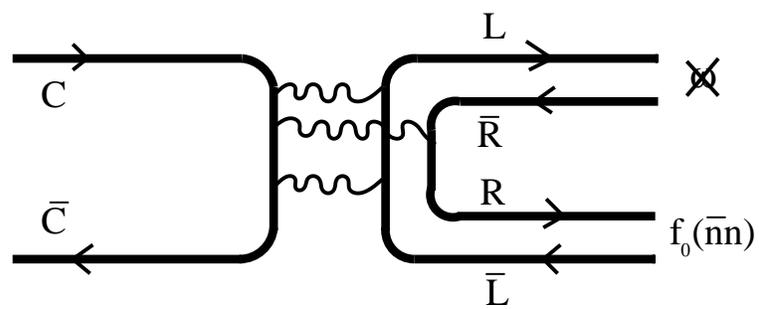}
\end{center}
\caption{The OZI-favoured decay which is forbidden by chiral symmetry.}
 \end{figure}

\begin{figure}
\begin{center}
\includegraphics*[width=6cm]{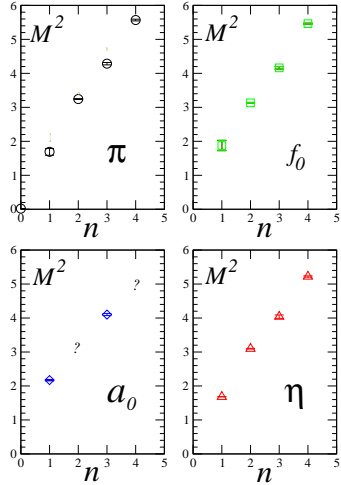}
\end{center}
\caption{Radial Regge trajectories for the four
successive high-lying J=0 $\pi, a_0, \bar n n ~~f_0$ and $\bar n n ~~\eta$
mesons. For details see Ref. \cite{G2} }
 \end{figure}

In particular, this selection rule should be taken into account
once a search of the two missing $a_0$-mesons, see Fig. 5, as
well as a confirmation of the $a_0$-meson around 2 GeV, are
undertaken. The existence of these states follows from the existence
of their chiral partners \cite{G2}. Namely, their 
$SU(2)_L \times SU(2)_R$ partners are the $\bar n n$ $\eta$-mesons,
while the $U(1)_A$ partners are the $\pi$-mesons. Hence 
masses of the high-lying $a_0$-mesons can be approximately
estimated from   the observed $\pi$- and $\eta$-mesons.
A search of the missing $a_0$-mesons can be performed e.g. in
$\chi_{c0} \rightarrow a_0(980) + a_0(?)$. This decay is not
chirally suppressed. On the contrary, charmonium decay into two mesons,
with one of them being $\pi$ or $\rho$ and the second one being a high-lying
$a_0$-meson, should be chirally suppressed.

As a conclusion, we have demonstrated a simple and clear
selection rule of dynamical origin, dictated by chiral symmetry, which forbids
some OZI-allowed decay mechanism. Hence some of the states,
where chiral symmetry is approximately restored, can be observed
in the OZI-forbidden channels, where there are no chiral symmetry
constraints. It is an important feature that this explanation
is coherent with the previously discussed restoration of chiral 
symmetry in excited hadrons.

\bigskip
\bigskip

The work was supported by the 
P16823-N08 project of the Austrian Science Fund.

\end{document}